# PRIVATELY POLICING DARK PATTERNS

*Gregory M. Dickinson*[*]

*Lawmakers around the country are crafting new laws to target "dark patterns"—user interface designs that trick or coerce users into enabling cell phone location tracking, sharing browsing data, initiating automatic billing, or making whatever other choices their designers prefer. Dark patterns pose a serious problem. In their most aggressive forms, they interfere with human autonomy, undermine customers' evaluation and selection of products, and distort online markets for goods and services. Yet crafting legislation is a major challenge: Persuasion and deception are difficult to distinguish, and shifting tech trends present an ever-moving target. To address these challenges, this Article proposes leveraging state private law to define and track dark patterns as they evolve. Judge-crafted decisional law can respond quickly to new techniques, flexibly define the boundary between permissible and impermissible designs, and bolster state and federal regulatory enforcement efforts by quickly identifying those designs that most undermine user autonomy.*

* Assistant Professor of Law and, by courtesy, Computer Science, St. Thomas University Benjamin L. Crump College of Law; Nonresidential Fellow, Stanford Law School, Program in Law, Science & Technology; J.D., Harvard Law School. I am grateful to the Program on Economics and Privacy at Antonin Scalia Law School, which provided funding for this research. For their insights and generous comments, thanks to participants at the George Mason University Law & Economics Center Research Roundtable on Regulating Privacy in December 2022, the New Voices in Internet & Computer Law Panel of the AALS Annual Meeting in January 2023, and the Nebraska Governance and Technology Center Law & Technology Workshop in March 2023. Finally, a special thank you to Laura Pedre, who provided excellent research assistance, and to Jesse (Tripp) Keefe, Savannah Grant, Millie Price, and others with the Georgia Law Review whose tireless energy and perceptive edits greatly improved this Article.

TABLE OF CONTENTS

# I. Introduction

There is a growing call for legislative action against "dark patterns"—user-interface designs that influence or even coerce users into making choices that are against their own interest, but beneficial to the website or app designer. Dark patterns can be used, for example, to trick users into sharing their personal data, into initiating recurring subscription payments for unwanted services, or into involuntarily submitting favorable product reviews. Such abuses have prompted widespread consumer complaints and alarmed regulators and legislators around the country who are concerned that dark patterns may harm the competitive market for online products and services by interfering with customers' purchasing decisions.

So far, however, there is little agreement about how to confront the challenge. At the federal level, the Federal Trade Commission (FTC) has taken the lead on the problem pursuant to its authority to police unfair and deceptive commercial practices. Some scholars have urged it to take on a greater role through fresh rulemaking or antitrust reform. In Congress, the leading proposal is the DETOUR Act, which would, among other things, broadly bar companies from employing user interfaces "with the purpose or substantial effect" of impairing "user autonomy, decision-making or choice."[1] Finally, leading the way at the state level are California and Colorado, which have recently enacted consumer privacy laws that restrict how companies obtain and use consumer data and expressly bar the use of "dark patterns" to obtain user consent.

All of these approaches, however, face similar obstacles: User-interface designs are so varied and evolve so quickly that it is difficult for lawmakers to define what designs should be prohibited; moreover, it is notoriously difficult to distinguish legitimate persuasion of customers from unlawful coercion, as evidenced by the large bodies of false advertising and unfair competition law devoted to the question; and, perhaps most importantly, state and federal enforcement resources are so limited that even if lawmakers were able to define precisely what sorts of user-interface designs should be prohibited, regulatory enforcers would still be able to focus their attention on only the very worst offenders.

---

[1] Deceptive Experiences to Online Users Reduction Act, S. 3330 and H.R. 6083, 117th Cong. § 3(a)(1) (2021) [hereinafter DETOUR Act].

Recognizing these difficulties, this Article takes a step back from the current spate of legislative and regulatory proposals to suggest a different route. Missing from the conversation to date has been the potential role of private law—that branch of law, including tort and contract law, that offers private rights of action for individuals to address grievances with one another. Private law's absence from the discussion is striking given that its basic features make it especially well-suited for combatting dark patterns.

First, private law is grounded in the decisional law of courts, which shape the law as they adjudicate disputes on a case-by-case-basis. Unlike public statutory or regulatory law, private law is eminently flexible and could quickly evolve to track new dark patterns as they emerge. Second, private law claims are litigated and funded by individuals. Individuals deceived by dark patterns have a strong incentive to pursue violators and also direct knowledge of the relevant facts. And, because private law actions are self-funded, a private-law approach to dark patterns could help to overcome the resource constraints that currently limit FTC enforcement efforts.

Given the inherent challenges of crafting dark-pattern-specific laws and the comparative advantages of private law, this Article suggests that lawmakers resist the impulse to target dark patterns directly and instead look for ways to harness and amplify the power of private law. Lawmakers might, for example, provide for statutory damages or attorney's fees where claimants successfully assert a tort or contract claim based on a defendant's use of a dark pattern. Or legislatures could create a completely new cause of action for harms caused by dark patterns.

The point is not to push for any particular approach; many could work, and a detailed comparison is beyond the scope of this Article. Instead, the aim is to illustrate how dark patterns are amenable to a private-law solution. Shifting enforcement to private litigants would direct fresh resources to policing dark patterns and press the law to evolve quickly to track new deceptive technologies as they appear.

## II. DARK PATTERNS

### A. WHAT IS A DARK PATTERN ANYWAY?

"Dark pattern" is a trendy phrase to describe a familiar phenomenon: website and smartphone app designs that encourage users to make choices against their best interest.[2] For example, an app asking a user to join its membership program (really, its email marketing list) might present a giant green "Opt In and Save $$$" button above a much smaller "No Thanks, I Don't Like Saving Money" option. Maybe there is no option to decline at all, just a barely noticeable "X" in a far-away corner. Or perhaps, having once joined the program, a user who changes her mind can only opt out by clicking through to a second and third screen, each confirming that she is really, really sure of her decision to leave. These insistent, confusing, and deceptive techniques were once the province mainly of unscrupulous sellers on fly-by-night websites.

In the internet's early days, what are now called dark patterns appeared most commonly in the form of the pop-up ad and its loathsome sibling, the pop-under. The ad might have told you, for example, that you were the website's one-millionth visitor and that you should follow the flashing link to claim your free iPod. Fine print would later reveal a random lottery drawing for the iPod to which you could add your name by submitting your personal information and agreeing to receive email marketing messages. Techniques like this were annoying, but manageable inconveniences; they were persistent, deceptive, and infuriating, but avoidable through browser settings, ad-blocking software, and by sticking to classy online neighborhoods.

---

[2] *See* Christoph Bösch, Benhamin Erb, Frank Kargl, Henning Kopp & Stefan Pfattheicher, *Tales from the Dark Side: Privacy Dark Strategies and Privacy Dark Patterns*, 4. PROC. PRIV. ENHANCING TECH. 243, 243 (2016) (describing dark patterns as "common building blocks that are used by service providers to deceive and mislead their users"); Colin M. Gray, Yuobo Kou, Bryan Battles, Joseph Hoggatt & Austin L. Toombs, *The Dark (Patterns) Side of UX Design*, PROC. 2018 CHI CONF. HUM. FACTORS COMPUTING SYS., Paper 534, at 1 (defining dark patterns as "instances where designers use their knowledge of human behavior (e.g., psychology) and the desires of end users to implement deceptive functionality that is not in the user's best interest"); Eric Ravenscraft, *How to Spot—and Avoid—Dark Patterns on the Web*, WIRED (July 29, 2020, 9:00 AM), https://www.wired.com/story/how-to-spot-avoid-dark-patterns/ (defining dark patterns as ways that software "can subtly trick users into doing things they didn't mean to").

Today's dark patterns are different. They are less obvious, more prevalent, and common even during routine, above-board transactions with well-known businesses. For example, last month I was offered the following choice during the final stage of Delta Air Lines' "Express Checkout" process:

**Figure 1. The trip protection offer seen during Delta Air Lines' "Express Checkout" process.[3]**

There is nothing wrong with Delta's offering to sell travel insurance with its flights. Some customers will want insurance and appreciate the convenience of purchasing it together with the flight in a single transaction. By structuring the insurance as an add-on, instead of including it in the price of the flight, Delta is able to

---

[3]    *See, e.g.*, *Trip Protection*, DELTA, https://www.delta.com/merch/searchTripInsuranceAction.action (last visited Mar. 25, 2023) (describing Delta's trip protection policy and benefits).

charge lower base flight prices while also offering insurance to those customers who want it.[4]

Observe, however, how the offer is presented. Note first that the page *requires* the customer to make a selection. The interface presents a radio button[5] that forces the customer to select either yes or no to continue. There is no way for the customer to ignore Delta's offer of optional insurance and simply proceed with the flight purchase. Much more so than would a separate section of optional, checkbox add-ons, this structure forces the customer to expend time and mental energy to evaluate the offer even if what she really wants is to focus on the flight purchase.

Next, having forced the customer to engage with the insurance offer, the page includes various "nudges"[6] that push the user toward Delta's preferred choice, the extra purchase. In the top right corner of the page is a banner indicating that trip-protection insurance is "Highly Recommended" (by whom is left unsaid). The bright-green color of both the banner and the "Yes" option suggest to the uncertain user that she should consider trip-protection insurance to be the recommended choice. The phrasing of the "Yes" and "No" options is also notable. By offering the options "Yes, protect my trip for $40.31 total" and "No, do not protect my $597.20 trip," the choice structure suggests an overly simple arithmetic comparison—that by paying only $40 the customer will be saving almost $600.[7] Moreover, the phrasing of the "No" option is designed to make the customer feel foolish by forcing her to select "No, do not protect my trip" instead of a simple "No, thank you." Finally, at the bottom of the page, Delta discloses that in the last three days 44,128 other customers have purchased trip-protection insurance (the count of those who declined the purchase is not shared) and that, according to Forbes.com, "purchasing travel insurance should be commonplace."[8]

---

[4] *See* Anthony K. Tjan, *The Pros and Cons of Bundled Pricing*, HARV. BUS. REV. (Feb. 26, 2010), https://hbr.org/2010/02/the-pros-and-cons-of-bundled-p (suggesting that "unbundling or a la carte pricing benefits the buyer" because "unbundled pricing creates transparency and allows you to pick exactly the options you want").

[5] *See Radio Buttons*, U.S. WEB DESIGN SYS., https://designsystem.digital.gov/components/radio-buttons/ (last visited Feb. 16, 2023) ("Radio buttons are a common way to allow users to make a single selection from a list of options. . . . [O]nly one radio button can be selected at a time . . . .").

[6] Transparent and, typically, well-meaning dark patterns are sometimes called "nudges." A voluminous literature has developed around the topic. *See* RICHARD H. THALER & CASS R. SUNSTEIN, NUDGE: THE FINAL EDITION 8 (revised ed. 2021) (defining a nudge as "any aspect of the choice architecture that alters people's behavior in a predictable way without forbidding any options or significantly changing their economic incentives"); Cass R. Sunstein, *Nudges Do Not Undermine Human Agency*, 38 J. CONSUMER POL'Y 207, 210 (2015) (reasoning that "[d]esirable nudges should undermine neither autonomy nor welfare" and that "with appropriate nudges, neither human agency nor consumer freedom is at risk"); Cass R Sunstein, *Do People Like Nudges?*, 68 ADMIN. L. REV. 177, 223 (2016) (concluding from a nationwide survey that "people will not, in general, rebel against nudges" because most nudges are transparent); CASS R. SUNSTEIN, HUMAN AGENCY AND BEHAVIORAL ECONOMICS: NUDGING FAST AND SLOW 1 (John Tomer ed., 2017) ("Nudges are specifically designed to preserve both agency and control. While nudges steer people in particular directions, they permit you to go your own way.").

Adding to the persuasive power of its design is the information that the page does *not* disclose: What happens to a customer whose flight is cancelled or whose baggage is mishandled but who has not purchased insurance? Only by searching on other areas of Delta's website does one learn that "if there is a significant delay, cancellation or schedule change," the customer is entitled to "a refund to your original form of payment"[9] and that customers whose bags are delayed are entitled to a rebate of any checked-bag fee and reimbursement of reasonable expenses incurred as a result of the delay, "generally determined as $50 USD per day."[10] In other words, the relevant comparison is not between complete compensation and complete loss of one's flight and baggage value, but between the reimbursement available under Delta's default policy versus its beefed-up trip-insurance policy (accounting also for the cost of the insurance and the probability of the loss occurring). That is a very different choice from the one that many consumers are likely to infer is on offer.

---

[7] *See supra* Figure 1. That is true, of course, only if the customer's trip is actually spoiled and she needs to invoke the insurance policy. The more appropriate comparison would be between the cost of the trip-protection insurance and the cost of the trip multiplied by the probability of it being interrupted in a way that is covered by the policy. The comparison Delta suggests is appropriate only where there is a 100% probability of trip interruption, which one hopes is not Delta's expectation.

[8] *Id.*

[9] *Frequently Asked Questions*, *Travel Planning Center*, DELTA, https://www.delta.com/us/en/travel-planning-center/travel-planning-faqs (last visited Feb. 26, 2023).

[10] *Delayed, Lost or Damaged Baggage*, DELTA, https://www.delta.com/us/en/baggage/delayed-lost-damaged-baggage (last visited Feb. 26, 2023).

The point is not to criticize Delta's travel-protection insurance sales. There is nothing wrong with offering add-on purchases or with presenting such offers in appealing terms. Instead, the point is to illustrate the phenomenon of "dark patterns" and highlight two of their key characteristics. First, dark patterns' "nudges" can be both subtle and powerful—especially for customers who are rushed or who do not communicate well in written English. The most powerful dark patterns look less like flashing pop-up ads of the 1990s and more like a crafty used-car saleswoman trying to get you in a car that you like at a price that *she* prefers. Second, dark patterns are everywhere,[11] influencing consumer purchases of everything from vacation packages[12] and wine clubs[13] to ink cartridges[14] and home thermostat systems.[15] It is impossible to navigate today's apps and websites without routinely encountering counterintuitive default settings, pressured choices, and deceptively structured interfaces, all designed to benefit their creators at the expense of the customer.

---

[11] User-interface specialist Harry Brignull maintains a lengthy and detailed catalog of examples. *See Hall of Shame*, DECEPTIVE DESIGN, https://www.deceptive.design/hall-of-shame/all (last visited Mar. 26, 2023) (including examples of dark patterns from companies such as Duolingo, Skype, DoorDash, and Slack).

[12] *See r/darkpatterns*, REDDIT (Oct. 26, 2020, 6:35 PM), https://www.reddit.com/r/darkpatterns/comments/jiozx6/bookingcom_pretends_that_it_has_a_working_button/ (recalling the non-functioning account deletion option by a user on the booking.com vacation portal, where it purports to cancel account and send email confirmation but does not).

[13] *See* (@darkpatterns), TWITTER (Feb. 21, 2021, 4:53 AM), https://twitter.com/darkpatterns/status/1363426610494517248 (showing WSJWine Advantage Club's concealed default recurring subscription).

[14] *See* (@darkpatterns), TWITTER (Aug. 15, 2022, 2:03 AM), https://twitter.com/darkpatterns/status/1559058241748967425 (describing the interface by which a customer canceling "HP Instant Ink" service is required to agree that "cartridges will no longer work after my final billing cycle ends, *even if they are already installed in my printer*" (emphasis added)).

[15] *See* Phil Hill (@PhilOnEdTech), TWITTER (Aug. 13, 2022, 11:42 AM), https://twitter.com/PhilOnEdTech/status/1558479009603657728 (providing a tweet by a customer who was locked out of his internet-connected home smart-thermostat system because the credit card on file with Amazon.com expired).

B. PROBLEMS WITH DARK PATTERNS

Dark patterns may be everywhere, but most are far less dangerous than the term might lead readers to believe. At their gentlest, there is hardly anything "dark" about them at all. Despite the scary name and media hullabaloo,[16] the majority of so-called dark patterns are just digital version of everyday social and business dealings: We present ourselves to the world in the best light, pitch our services in compelling terms, design our product lines to our companies' long-term advantage, and exercise every art of persuasion to make and keep the sale. What could be more American than that? Self-interested business dealings are as old as humanity, and even if courts or legislatures wanted to regulate them, they would quickly bump against competing interests in freedom of speech and contract.

The problem comes when dark patterns cross that invisible line that separates authentic persuasion from deception and coercion. Attractive ads full of beautiful, smiling models and well-behaved children are par for the course. So, too, are high-pressure sales pitches for used cars, time-shares, or, in the digital world, add-on insurance products accompanying airfare purchases.[17] These tactics lie toward the lighter end of the dark-pattern spectrum—aggressive and annoying, but also transparent and navigable. Users can self-police the problem by taking their business elsewhere. The various interests at stake, including free speech, freedom of contract, product differentiation, market efficiency, and consumer peace of mind are managed by consumers' avoiding or "naming and shaming"[18] sellers who employ undesirable tactics.

---

[16] *See, e.g.*, Emily Stewart, *The Psychological Traps of Online Shopping, Explained*, Vox (Dec. 15, 2022, 8:30 AM), https://www.vox.com/the-goods/23505330/online-shopping-ecomerce-tricks-dark-patterns-deceptive-design (warning of dark patterns tricking consumers); Jennifer Valentino-DeVries, *How E-Commerce Sites Manipulate You into Buying Things You May Not Want*, N.Y. Times (June 24, 2019), https://www.nytimes.com/2019/06/24/technology/e-commerce-dark-patterns-psychology.html (describing the methods companies use to encourage consumers to make purchases); Catherine Thorbecke, *How Companies Subtly Trick Users Online with "Dark Patterns,"* CNN: Bus. (July 18, 2022, 11:32 AM), https://www.cnn.com/2022/07/16/tech/dark-patterns-what-to-know/index.html (same).

[17] *See* discussion *supra* section II.A.

[18] Shaming companies that employ dark pattern into better behavior has been the aim of Harry Brignull, who coined the phrase "dark patterns" and operates the Deceptive Design

At the other end of the spectrum lie those patterns that, in contrast with mere persuasion, employ techniques that are so deceptive or coercive as to undermine the users' free choices. For analogue-world examples, think of common-law fraud, blackmail,[19] or quid pro quo sexual harassment. Each involves a "choice," but the offer is conveyed in false or deceptive terms or with unlawful conditions attached that undermine consent. Dark patterns sometimes employ similar tactics, which interfere with users' decision-making processes and require a legal remedy notwithstanding the user's apparent "choice."

*1. Unlawful Bargains.* In the extreme examples of blackmail, extortion, quid pro quo sexual harassment, and similar abuses of power, the supposed bargain is trumped by society's collective decision that certain conduct is simply out of bounds. Consider the recent spate of payments extorted from hospitals through ransomware.[20] In a typical scenario, attackers break into a hospital's computer network and encrypt its patient-record databases, which renders them inoperable without the encryption key. The attackers then offer to decrypt the data in exchange for a large payment in digital currency.[21] Even assuming the offer is truthful and the attackers will decrypt the data after receiving payment, the offer is problematic because the attackers have no right to the benefit they are offering for sale. Whether in person with a gun or virtually by dark pattern, options like "hand over the money, or else . . ." are simply not choices that can be lawfully offered.

*2. Invalid Consent.* A much more common problem with dark-pattern-induced transactions is their potential interference with individual autonomy. A choice can be offered in such deceptive terms that a user should not be bound by her decision despite her

---

website to collect examples of dark patterns submitted by users. *See* DECEPTIVE DESIGN, https://www.deceptive.design (last visited Feb. 19, 2023).

[19] *See generally* James Lindgren, *Unraveling the Paradox of Blackmail*, 84 COLUM. L. REV. 670, 670 (1984) ("In blackmail, the heart of the problem is that two separate acts, each of which is a moral and legal right, can combine to make a moral and legal wrong.").

[20] *See* Robert McMillian & Jenny Strasburg, *Mounting Ransomware Attacks Morph into a Deadly Concern*, WALL ST. J. (Sept. 30, 2020, 12:39 PM), https://www.wsj.com/articles/mounting-ransomware-attacks-morph-into-a-deadly-concern-11601483945 (discussing the phenomenon of ransomware attacks on hospitals and the medical industry more broadly).

[21] *Id.*

apparent consent to the offered terms. Consider cases of fraud.[22] In *Federal Trade Commission v. Effen Ads, LLC*,[23] for example, the Federal Trade Commission initiated an action against Effen Ads, alleging that the company sent consumers unsolicited email messages that falsely purported to originate from reputable news outlets,[24] hosted fake online news stories mimicking Fox News, CNN, and other websites,[25] and then presented visitors with ads for work-from-home schemes.[26] Effen Ads promised visitors they could make "significant income with little effort working from home" if they paid an up-front fee of ninety-seven dollars.[27] In reality, despite as assurances of "A Certified, Proven And Guaranteed Home Based Business Jobs [sic] To Make \$379/Day From Home,"[28] those who paid the fee did not receive jobs or even links to job postings.[29] After payment, users received access to a "Member's Area" of the website that contained online videos and tutorials about how they could start their own online "link-posting" businesses.[30]

The outright lies that Effen Ads' website allegedly presented to its users are especially egregious examples of what is called "sneaking"—a dark pattern of concealing or delaying disclosure of important information from the user until it is too late or too burdensome to change course.[31] The website obscured the true origin of the news stories that lured visitors to the site and then delayed revealing the true nature of the service being offered—a collection of videos and tutorials—until after users had already handed over the money.[32] The *Effen Ads* case[33] shows one way that dark patterns can be so deceptive as to undermine user consent to the transaction and justify legal intervention.[34]

---

[22] A party induced to act by another's intentional "material misrepresentation of fact, opinion, intention, or law," has numerous remedies available to avoid payment or recover losses, including rescission of the transaction, suit for fraud or breach of contract, or fraudulent inducement as a defense to a demand for payment. *See* RESTATEMENT (THIRD) OF TORTS: LIAB. FOR ECON. HARM § 9 (AM. L. INST. 2020) (explaining the relationship between common-law fraud and similar concepts in the laws of restitution and contract, providing that "[o]ne who fraudulently makes a material misrepresentation of fact, opinion, intention, or law, for the purpose of inducing another to act or refrain from acting, is subject to liability for economic loss caused by the other's justifiable reliance on the misrepresentation").

[23] Compl. ¶ 2, Fed. Trade Comm'n v. Effen Ads, LLC, No. 2:19-CV-00945 (D. Utah Nov. 26, 2019) [hereinafter Effen Ads Complaint].

[24] *Id.* ¶¶ 53–54.

[25] *Id.* ¶¶ 82–103.

[26] *Id.* ¶¶ 3, 79

[27] *Id.* ¶¶ 3–8, 61–81.

[28] *Id.* ¶ 75.

[29] *Id.* ¶ 79.

[30] *Id.* ¶ 75.

[31] *See* Gray et al., *supra* note 2, at 6 (defining sneaking as "an attempt to hide, disguise, or delay the divulging of information that has relevance to the user").

But the problem is much broader than purveyors of shady work-from-home schemes. Internet users interact with confusing, not-quite-transparent apps and websites on a daily basis. Especially bold have been the data-hungry apps of the nation's tech sector. Modern apps and services often require vast troves of customer data—data that, under emerging state data privacy laws, cannot be collected and used without the consent of the user. Companies trying to persuade their customers to release their data have naturally turned to the most persuasive interfaces possible, often including dark patterns.

Another common problem area is the burgeoning online market for subscription services. Today one can find a subscription service for almost anything—the usual slate of streaming services with Netflix, Hulu, or Disney+, stylist-selected clothing with the likes of StitchFix and Wantable, monthly shipments of bamboo-made toilet paper with Cloud Paper, for a "forest-friendly flush,"[35] or, for the backyard poultry farming enthusiast, the Henny+Roo monthly Chicken Keepers Box.[36] Even auto manufacturers are getting into the subscription business, with monthly fees to enable top acceleration and to keep heated seats working.[37]

---

[32] *See* Effen Ads Complaint, *supra* note 23, ¶¶ 3–8, 61–81.

[33] Effen Ads ultimately stipulated to an order permanently enjoining it from selling its business-coaching services and requiring it to pay more than eleven million dollars in damages. Stipulated Order at 8–15, Fed. Trade Comm'n v. Effen Ads, LLC, No. 2:19-CV-00945 (D. Utah Dec. 20, 2019).

[34] *See* Justin (Gus) Hurwitz, *Designing a Pattern, Darkly*, 22 N.C. J. L. & TECH. 57, 57 (2020) (expressing skepticism of the need for legal change, but agreeing that in their strong versions, dark patterns undermine consent and warrant legal action)

[35] *See* CLOUD PAPER, https://cloudpaper.co/ (last visited on Jan. 30, 2023).

[36] *See* HENRY AND ROO, https://hennyandroo.cratejoy.com/ (last visited on Jan. 30, 2023).

[37] *See* Tom Gerken, *Mercedes-Benz to Introduce Acceleration Subscription Fee*, BBC (Nov. 24, 2022), https://www.bbc.com/news/technology-63743597 (noting that Mercedes-Benz is introducing a $1,200 annual subscription fee for acceleration); Allison Prang, *Want Heated Seats in Your BMW? There's a Monthly Fee for That*, WALL ST. J. (July 13, 2022), https://

There is nothing wrong either with subscription-payment arrangements or with the unusual services some customers apparently prefer. The problem comes when the true terms of the arrangement are obscured. Companies selling subscription services might conceal the fact that, after an initial trial period, the service will begin auto billing the customer's account.[38] Or, having once attracted a customer, a seller might make it difficult for the user to cancel or change the frequency of the subscription to keep the user locked-in to the arrangement. Such dark-pattern-ladened sign-up and cancellation processes so undermine consent as to justify a legal remedy to unwind the transaction.

Consumer deception through dark patterns is frighteningly common, and also easier than one might think. Recent experimental work by Jamie Luguri and Lior Jacob Strahilevitz suggests that even mild dark patterns can double the percentage of consumers who agree to pay for dubious services and that aggressive dark patterns might quadruple the percentage.[39] The FTC has accordingly warned[40] that it intends to increase its enforcement efforts against companies that employ dark patterns to trick consumers into paying for services.[41] Of course, when exactly a

---

www.wsj.com/articles/want-heated-seats-in-your-bmw-theres-a-monthly-fee-for-that-11657719715 (reporting that BMW is introducing subscription fees for heated seats).

[38] *See* Enforcement Policy Statement Regarding Negative Option Marketing Notice, 86 Fed. Reg. 60822 (Nov. 4, 2021) [hereinafter FTC Policy on Negative Option Marketing Notice] (discussing the application of various laws to "negative option marketing", or the conditioning of a term or condition under which the seller may interpret a consumer's silence or failure to take affirmative action to reject a good or service or to cancel the agreement as acceptance or continuing acceptance of the offer").

[39] Jamie Luguri & Lior Jacob Strahilevitz, *Shining a Light on Dark Patterns*, 13 J. LEGAL ANALYSIS 43, 46 (2021).

[40] *See* Press Release, Fed. Trade Comm'n, FTC to Ramp up Enforcement Against Illegal Dark Patterns that Trick or Trap Consumers into Subscriptions (Oct. 28. 2021), https://www.ftc.gov/news-events/news/press-releases/2021/10/ftc-ramp-enforcement-against-illegal-dark-patterns-trick-or-trap-consumers-subscriptions (announcing the FTC's new enforcement policy regarding negative option marketing); FED. TRADE COMM'N, BRINGING DARK PATTERNS TO LIGHT 1–3 (2022) [hereinafter BRINGING DARK PATTERNS TO LIGHT], https://www.ftc.gov/system/files/ftc_gov/pdf/P214800%20Dark%20Patterns%20Report%209.14.2022%20-%20FINAL.pdf (describing various types of dark patterns, explaining how they affect consumers, and discussing FTC efforts to combat them).

[41] Invalid consent mechanisms can also affect noneconomic interests such as consumer privacy. *See* Jessica Rich, *Five Reforms the FTC Can Undertake Now to Strengthen the*

carefully phrased ad or strategic interface design crosses the line[42] to invalidate consent is an incredibly difficult question that is taken up later in the discussion.[43] For now, the key point is that certain dark pattern designs interfere with consumer consent to such a degree as to undermine consent and warrant legal intervention.

*3. Market Inefficiency.* Dark patterns also interfere with market efficiency, a result that flows directly from their potential to undermine user consent. The same dark-pattern strategies that undermine consent also interfere with other aspects of market participants' decision-making processes. Most obvious are those transactions that would not be entered into absent the deception or coercion of some dark pattern. *Effen Ads* serves again as a ready example. Visitors to the company's website would never have paid it ninety-seven dollars were it not for the apparent offer of an easy, work-from-home job. Had the website made clear the very different product that was actually being offered, its customers would almost certainly have declined the "opportunity."

Less extreme misstatements regarding a product or service might extract a higher price from the customer, rather than induce an entirely unwanted transaction. Or, more subtly, dark patterns might encourage users to choose higher-priced products or to select privacy settings that funnel valuable personal information to the

---

*Agency*, BROOKINGS (Mar. 1, 2021) (observing that "over the last two decades, the FTC has taken action against a wide range of harms that are non-economic in nature," including various privacy violations, and urging the FTC to issue a new policy statement expanding its concept of consumer harm); *see also* WOODROW HARTZOG, PRIVACY'S BLUEPRINT (2018) (discussing the privacy implications of software design choices and the failures of the current model of privacy law).

[42] Similar problems crop up all throughout the law from contract formation and fraudulent inducement to advertising puffery and "dealer talk." *See, e.g.*, Wei Wen, *A Comparative Analysis of Sino-American Contractual Writing Attributes: Underpinnings for China's Future Uniform Civil Code to Mandate Writing for Land Sale Contracts*, 16 S.C. J. INT'L L. & BUS. 23, 30 (2020) (discussing line-drawing problems related to contract formation and fraud); David A. Hoffman, *The Best Puffery Article Ever*, 91 IOWA L. REV. 1395, 1398–99 (2006) (discussing line-drawing problems related to advertising puffery).

[43] *See infra* Parts III–IV; *see also* Neil Richards & Woodrow Hartzog, *The Pathologies of Digital Consent*, 96 WASH. U. L. REV. 1461, 1489 (2019) (describing several examples of what the authors call "coerced consent"—including when a user does not have the option to decline but only to accept "later," or a user interface that words the option to decline in such a way as to shame the user into compliance—which at scale, the authors argue, can accumulate to deplete a user's resolve with respect to their privacy choices).

application's creator.[44] One prominent example is Facebook's initial slate of default privacy settings, which were set, among other things, to make users' posts publicly searchable on Google and to enable geolocation services to track the users' location—settings users might have configured differently had they been presented a clear choice, and which allowed Facebook to collect commercially valuable data on its users.[45]

The extent of dark patterns' interference with customer decisions, the line separating persuasion from deception, the extent to which legal intervention is required or justified, and whether existing legal regimes adequately cabin the anticompetitive effects of dark patterns are all hotly debated questions.[46] The point at this stage is to show dark patterns' potential to distort markets.

---

[44] Gregory Day & Abbey Stemler, *Are Dark Patterns Anticompetetive?*, 72 ALA. L. REV. 1, 36–39 (2020) (theorizing that tech firms with significant market power might cause noneconomic harms, such as extracting user time, attention, and data despite offering inferior quality products).

[45] *See, e.g.*, Alex Hern, *Facebook Personal Data Use and Privacy Settings Ruled Illegal by German Court*, THE GUARDIAN (Feb. 12, 2018, 10:28 A.M.) https://www.theguardian.com/technology/2018/feb/12/facebook-personal-data-privacy-settings-ruled-illegal-german-court (describing a recent suit ruling that those default privacy settings were illegal in Germany); Natasha Lomas, *Antitrust Case Against Facebook's "Super-Profiling" Back on Track After German Court Ruling*, TECHCRUNCH (June 23, 2020, 12:51 PM) https://techcrunch.com/2020/06/23/antitrust-case-against-facebooks-super-profiling-back-on-track-after-german-federal-court-ruling/ (describing a German court ruling that default, prechecked privacy settings did not constitute valid consent under European Union privacy law); David Meyer, *Facebook Loses Germany Court Battle Over Privacy Settings*, CBS NEWS (Feb. 13, 2018), https://www.cbsnews.com/news/germany-facebook-court-case-privacy-settings-terms-of-use-brought-vzbz (same).

[46] *See, e.g.*, James C. Cooper & John M. Yun, *Antitrust & Privacy: It's Complicated*, 2022 J.L. TECH. & POL'Y 343, 369–92 (finding no relationship between website and app privacy protections and market concentration and concluding that suboptimal privacy protections are more likely attributable to informational problems with customers access to and reliance on information about companies privacy practices); Day & Stemler, *supra* note 44, at 36–39 (attributing privacy and other noneconomic consumer harms attributable to tech companies' concentrated market power); Hurwitz, *supra* note 34, at 95 (observing the problems raised by dark patterns but concluding that because most dark patterns "involve making representations or engaging in practices that are designed to deceive consumer," they "should fall within the ambit of the FTC's consumer protection authority"); John M. Newman, *Antitrust in Zero-Price Markets: Foundations*, 164 U. PA. L. REV. 149, 203–06 (2015) (suggesting a role for antitrust law to protect consumer privacy on the ground that in zero-price markets "information (instead of money) is the relevant currency").

\*       \*       \*

This initial consideration of dark patterns has revealed a few key points: Dark patterns are not a new phenomenon, but instead the application of age-old human persuasion and trickery to the online world. They appear in a new, digital form and vary greatly in their mechanisms of persuasion and deception, yet their effects are familiar. At one end of the spectrum, they range from mildly to moderately bothersome, but they may also provide valuable information to consumers by, for example, introducing new offerings or helping to differentiate competing products or services. At the more extreme end of the spectrum, however, they undermine autonomous decision making, distorting online markets, and unlawfully coerce consumer behavior.

## III. REGULATORY AND STATUTORY APPROACHES

Dark patterns are a common source of customer frustration. Nobody likes feeling duped, and consumer complaints have naturally attracted the scrutiny of regulators and legislators.[47] To date, however, there is little consensus on what (if anything) should be done to address the issue. This Part presents a few potential responses and then proceeds to discuss the special challenges that dark patterns present lawmakers.

---

[47] *See* Valentino-Devries, *supra* note 16 (discussing consumers' growing frustration with dark patterns and potential legislative responses); Yoree Koh & Jessica Kuronen, *How Tech Giants Get You To Click This (and Not That)*, WALL ST. J. (May 31, 2019, 11:18 AM), https://www.wsj.com/articles/how-tech-giants-get-you-to-click-this-and-not-that-11559315900 (same); Thomas Germain, *New Dark Patterns Tip Line Lets You Report Manipulative Online Practices*, CONSUMER REPS. (May 19, 2021), https://www.consumerreports.org/digital-rights/dark-patterns-tip-line-report-manipulative-practices-a1196931056 (announcing launch of a dark-pattern tip line for consumers submit complaints); Lesley Fair, *$245 Million FTC Settlement Alleges Fortnite Owner Epic Games Used Digital Dark Patterns to Charge Players for Unwanted In-Game Purchases*, FED. TRADE COMM'N (Dec. 19, 2022), https://www.ftc.gov/business-guidance/blog/2022/12/245-million-ftc-settlement-alleges-fortnite-owner-epic-games-used-digital-dark-patterns-charge ("[Epic Games] received more than a million complaints about unwanted charges.").

A. THE FEDERAL TRADE COMMISSION

At the forefront has been the FTC, the leading enforcer of the nation's consumer protection laws.[48] The FTC's mandate comes not from any new, dark-pattern-specific law, but rather from its long-standing authorization under the FTC Act to police "unfair methods of competition" and "unfair or deceptive acts or practices affecting commerce."[49]

The FTC's mandate is exceptionally broad, encompassing everything from anticompetitive mergers and acquisitions to false advertising, identity theft, and data breaches. Yet despite its limited resources and expansive enforcement responsibilities, the FTC has chosen to devote special attention to the problem of dark patterns. In April 2021, the FTC convened a dark-pattern workshop, to which it invited members of Congress, researchers, legal experts, and others, to consider the risks that digital dark patterns pose to consumers and to develop strategies for combatting them.[50] Soon thereafter, the FTC issued a new policy statement on negative option marketing—a common dark-pattern tactic in which a consumer's silence or failure to take affirmative action to reject a good or service is interpreted as acceptance of that good or service.[51] Finally, in late 2022 the FTC issued a report on dark patterns, which categorizes patterns by type and analyzes how each may run afoul of the FTC Act's prohibitions on unfair or deceptive acts or practices.[52] By signaling its concerns with dark patterns, threatening enforcement actions against sellers who use them, and offering guidelines for compliance, the FTC hopes to use its existing powers under the FTC Act to discourage entities from using dark patterns.

---

[48] *See Protecting Consumers*, FED. TRADE COMM'N, https://www.ftc.gov/news-events/topics/truth-advertising/protecting-consumers (last visited Feb. 18, 2023) ("As the nation's consumer protection agency, the Federal Trade Commission has a broad mandate to protect consumers from fraud and deception in the marketplace."); The states, which have their own analogous consumer protection laws, have largely followed the FTC's lead, relying on existing law to combat dark patterns. Some states have opted to enact new laws specifically targeting dark patterns, however. *See* Catherine Zhu, *Dark Patterns—A New Frontier in Privacy Regulation*, REUTERS (July 29, 2021, 11:56 AM), https://www.reuters.com/legal/legalindustry/dark-patterns-new-frontier-privacy-regulation-2021-07-29/ ("Under Section 5 of the FTC Act, the FTC has the authority to prosecute companies for unfair or deceptive trade practices, which it has exercised against businesses for the use of dark patterns. . . . At the state level, both California and Colorado have passed consumer privacy legislation banning the use of dark patterns, and other states are contemplating doing the same.").

[49] FTC Act § 5(a), 15 U.S.C. § 45(a)(1).

[50] *See* BRINGING DARK PATTERNS TO LIGHT, *supra* note 40, at 4–6 (describing the rise of dark patterns and providing recommendations to companies).

[51] *See* Enforcement Policy Statement Regarding Negative Option Marketing, 86 Fed. Reg. 60822 (Oct. 29, 2021) (guiding businesses on the FTC's "interpretation of existing law as it applies to negative option practices").

There is much to commend in this approach. First, the FTC Act is already on the books, ready at hand. Combating dark patterns through existing prohibitions on "unfair" and "deceptive" methods of competition requires no new legislation or formal rulemaking. Second, the FTC Act's broad language dates from the early twentieth century, and has now been refined through decades of case law. Relying on the Act's time-tested formulation reduces the risk that Congress or the FTC might miscalculate and inadvertently create an overaggressive, market-disrupting rule were it to develop a new one from scratch. Finally, the FTC Act's language is more than up to the task. In their problematic forms, dark patterns are specifically designed to mislead consumers, conduct that falls squarely within FTC Act Section 5's prohibition on "unfair or deceptive acts or practices."[53]

## B. DIRECTLY ADDRESSING DARK PATTERNS

Beyond enforcement under the FTC Act, Congress is also considering new laws that would target dark patterns directly. And some states have already done so, notably California and Colorado, whose recently enacted privacy laws specify that dark-pattern-induced agreements do not satisfy statutory consent requirements regarding use of consumer data.

*1. The DETOUR Act.* At the national level, Congress has been considering legislation to build consumer data privacy protections into federal law. Most directly implicating dark patterns is the proposed DETOUR Act,[54] which would combat deceptive online practices in three ways: First, for children under the age of thirteen, online entities would be barred from designing user interfaces "with the purpose or substantial effect of causing . . . compulsive usage."[55] The Act specifically targets "video auto-play functions," like those employed by YouTube, Netflix, and other streaming services, but the language is not limited to those strategies. The proposed prohibition on designs that produce "compulsive usage" is well intentioned, but broad enough that it could capture all manner of gamified app designs,[56] even those whose addictive properties are an unintentional side effect of an app designed to be fun to use.

---

[52] BRINGING DARK PATTERNS TO LIGHT, *supra* note 40.

[53] *See* Hurwitz, *supra* note 34, at 95 (explaining how many dark patterns already fall under existing law and the "FTC's consumer protection authority"); FTC Act § 5(a), 15 U.S.C. § 45(a)(2) ("The Commission is hereby empowered and directed to prevent persons, partnerships, or corporations . . . from using unfair methods of competition in or affecting commerce and unfair or deceptive acts or practices in or affecting commerce.").

Second, for all users, regardless of age, the Act would prohibit companies from employing any user interface "with the purpose or substantial effect of obscuring, subverting, or impairing user autonomy, decision making, or choice to obtain consent or user data."[57] Strangely, this prohibition is at once both narrow and broad. The law is narrow in that it applies only where user autonomy is undermined "to obtain consent or user data," not for other purposes.[58] Yet it is also exceptionally broad in that it applies to all designs that have the "effect" of "obscuring, subverting, or impairing user autonomy or choice" without regard to context or the interface designer's mental state.[59] The bar applies whether or not the app or website designer intended or even knew that the interface design would be confusing to users. And it makes no provision for hard-sell-type advertising strategies that are permissible under long-standing law, but which could nonetheless be said to "subvert" user autonomy or choice.[60]

Third, the Act would bar the common practice of A/B interface testing, in which companies tweak interface designs by splitting users into groups and comparing usage metrics between groups.[61] Such testing is a mainstay of interface design because it helps companies optimize their user interface designs. For example, a company might split users into A and B test groups, present each a slightly different user interface, and measure which group is able to complete routine tasks more quickly, which group rates the app more highly, or which group makes fewer requests for assistance. The problem comes when the same techniques are employed to maximize user consent rates. Unlike A/B testing for ease of use, which produces navigable, streamlined interfaces, A/B testing to maximize consent rates tends to produce opaque interface designs that obscure the privacy implications of a user's selection. What better way to get a user's consent than to trick her into thinking her click means something else altogether? A/B testing has thus become a target for lawmakers, but the DETOUR Act's failure to include a safe harbor for less nefarious uses of the technique risks undermining interface designers' ability to develop user-friendly apps and websites.[62]

---

[54] Deceptive Experiences to Online Users Reduction Act, S. 3330 and H.R. 6083, 117th Cong. (2021).

[55] *Id.* § 3(a)(3).

[56] *Id.*

[57] *Id.* § 3(a)(1).

[58] *Id.*

[59] *Id.*

[60] *Id.*

[61] *Id.* § 3(a)(2).

*2. State Consumer Privacy Laws.* Legislation at the state level is further along. Both California[63] and Colorado[64] have adopted data privacy laws that restrict how companies obtain and use consumer data.[65] These laws grant consumers various rights over their personal data, including the right to demand that companies delete their personal information, the right to obtain a portable copy of their data, and the right to transparent notice of the types of data being collected about them. In some circumstances the laws also require entities to obtain a consumer's consent to the use of her data.[66]

Where consent is required, both California and Colorado expressly bar entities from using dark patterns. The laws provide that "agreement obtained through use of dark patterns does not constitute consent"[67] and define dark pattern to mean any user interface "with the substantial effect of subverting or impairing user autonomy."[68] Importantly, the California privacy law also directs the relevant state agency to more precisely define what constitutes a "dark pattern." so as to provide greater guidance to regulated parties.

---

[62] *See* Katharine Schwab, *We're All Being Manipulated by A/B Testing All the Time*, FAST CO. (Feb. 24, 2019), https://www.fastcompany.com/90306916/were-all-being-manipulated-by-a-b-testing-all-the-time (discussing the ethical issues regarding A/B testing).

[63] California Consumer Privacy Act of 2018, CAL. CIV. CODE § 1798.100 (West 2022).

[64] Colorado Consumer Privacy Act, COLO. REV. STAT. § 6-1-1313 (West 2022).

[65] The proposed federal DETOUR Act would take a similar approach. *See* discussion *supra* section III.B.*1*.

[66] *See* California Consumer Privacy Act of 2018 § 1798.135(b)(2), (c)(4) (requiring express re-consent to use of personal information if a consumer has previously opted out, and requiring express consent before an entity may share the personal information of a consumer under sixteen years of age and that the entity "wait for at least 12 months before requesting the consumer's consent again").

Exercising that rulemaking authority, in February 2023, the California Privacy Protection Agency issued new regulations to define dark patterns in more detail. The regulations provide that to support valid consent under the California Consumer Privacy Act (CCPA), a user interface must, among many other things: (1) "use language that is easy for consumers to read and understand"; (2) employ "symmetry in choice" so that the steps required to select a more privacy-protective option are not more numerous than the steps to select a less privacy-protective option; (3) avoid confusing language and interactive elements such as yes-no toggle buttons that lack sufficient descriptions or that are described using double negatives; (4) avoid "manipulative language or choice architecture" such as "wording that guilts or shames the user into making a particular choice"; and (5) "not add unnecessary burden or friction to the process."[69]

Such detailed regulations have obvious advantages. In contrast to the FTC Act's broad prohibition of "unfair" or "deceptive" practices,[70] California's regulations provide a detailed description of what constitutes a "dark pattern," along with specific examples, which help regulated entities comply with the requirements and give them less excuse for failing to do so.[71] And, unlike informal guidance documents, the regulations carry the force of law and require a formal rulemaking process to amend, giving parties greater assurance that the law will remain the same over time and across shifts in government administrations.[72]

---

[67] *Id.* § 1798.140(h); Colorado Privacy Act § 6-1-1303(5).

[68] California Consumer Privacy Act of 2018 § 1798.140(l); Colorado Privacy Act § 6-1-1303(9).

[69] *See* CAL. CODE REGS. tit. 11, § 7004(a)(1)–(5).

[70] *See* FTC Act § 5(a), 15 U.S.C. § 45(a)(1) ("Unfair methods of competition in or affecting commerce, and *unfair* or *deceptive* acts or practices in or affecting commerce, are hereby declared unlawful." (emphasis added)).

[71] *See* CAL. CODE REGS. tit. 11 § 7004(5)(b)–(c) (providing detailed regulations with examples of violative practices).

But the approach has its limitations. In a fast-moving field like computer-human interface design, the examples and specific prohibitions in the regulations will quickly fall out of date. Indeed, the regulations themselves are certain to spur a shift away from the specifically proscribed techniques and toward those that fall into the regulations' gray areas. Anticipating exactly that, the regulations include not only specific limitations, but also broad ones, such as the prohibition of designs that produce "unnecessary burden or friction,"[73] the "unintuitive placement of buttons," and "manipulative language or choice architecture."[74] Capacious catch-all provisions ensure that future technologies cannot slip through the cracks, but they also undermine the very specificity and predictability that are the regulations' strengths.

Finally, even putting to one side the difficulty of keeping tech-industry laws up-to-date, California's dark pattern regulations are severely limited in scope: The regulations were promulgated pursuant to statutory authority to define what consumer action counts as consent to collection or use of personal data in compliance with the CCPA.[75] But the problem with dark patterns is much broader than that. Dark patterns have the potential to undermine user consent in any online transaction, not just CCPA-mandated consents to use of personal data. A law targeting dark patterns would ideally address *all* instances where interface designs interfere with consent, not just those that occur within the CCPA-mandated disclosure and consent process.

---

[72] *See* CAL. GOV'T CODE § 11346–48 (detailing regular and emergency rulemaking procedures); Office of Administrative Law, *California Code of Regulations*, STATE OF CAL., https://oal.ca.gov/publications/ccr/ (last visited Feb. 15, 2023) ("Properly adopted regulations that have been filed with the Secretary of State have the force of law.").

[73] CAL. CODE REGS. tit. 11 § 7004(a)(5).

[74] *Id.* § 7004(a)(3)(C), (a)(4).

[75] *See* California Consumer Privacy Act of 2018, CAL. CIV. CODE § 1798.185 (West 2022) (authorizing rulemaking to "further the purposes" of the CCPA, "including, but not limited to" various enumerated areas, including the statute's prohibition of dark patterns); CAL. CODE REGS. tit. 11 § 7000 ("These regulations govern compliance with the California Consumer Privacy Act. . . .").

C. CHALLENGES OF DARK-PATTERN LAWMAKING

Whatever the path forward, whether federal privacy legislation, expanded antitrust enforcement, privacy-enhancing technology, or something else altogether, dark patterns present a knotty problem: Persuasion and coercion of consumers are notoriously difficult to distinguish; user interface designs change rapidly, and, to top it off, state and federal enforcement resources are so limited that regulators are able to focus only on the very worst offenders. This section introduces the difficulties inherent in dark-pattern lawmaking to prepare the way for discussion in Part IV of how private law can play a role in policing dark patterns.

*1. Evolving Technology.* As noted above,[76] one major obstacle to legislative or regulatory action is the pace of technological change. Think back to user interfaces of ten or even five years ago. Many are scarcely recognizable today. As app makers implement new features, streamline designs, and target new audiences, they must update the interfaces through which they communicate with users. And beyond updates to the content and functionality of apps and websites themselves, new design technologies regularly emerge that make the user experience ever-more immersive. Static web pages yielded to Adobe Flash, which yielded to HTML5, which itself will yield, or evolve, to augmented and virtual reality. Wherever human ingenuity leads, human huckstering and fresh dark patterns are sure to follow.

The problem for lawmakers is that the traditional legislative and regulatory processes are principally reactive. When a new problem arises, lawmakers respond by collecting data, deliberating to identify available options and form consensus, and, finally, by crafting legislation or regulations to address the problem.[77] Lawmakers, of course, do their best to anticipate future changes and craft legislation at an appropriate level of abstraction. But sometimes the future brings unforeseeable shifts in technology. To be truly future-proof, a dark-pattern law would need to say simply

---

[76] *See supra* section III.B (noting the problem in the context of California's Consumer Privacy Act regulations).

[77] *See* Wulf A. Kaal & Erik P.M. Vermeulen, *How to Regulate Disruptive Innovation—From Facts to Data*, 57 JURIMETRICS J. 169, 170–73 (2017) (discussing the traditional ex post rulemaking process and the challenge it faces when regulating rapidly evolving technology).

"do not be *too* tricky"—surely the right rule, but not one that offers much guidance to regulated entities.[78]

*2. Persuasion Versus Coercion.* Even if technology would ever sit still and let lawmakers regulate it, they would soon encounter another obstacle: It is really, really hard to draw a line between lawful persuasion and unlawful coercion. One woman's hard sell is another woman's extortion. The two are difficult enough to distinguish after the fact, with complete information about the parties and transaction in question. The feat is harder still for legislators and regulators trying to define the boundary ex ante, for all parties, all times, and all circumstances, without knowledge of the specific facts.

Consider Congress's effort with the Federal Trade Commission Act,[79] the statute that created the FTC and empowers it to police everything from anticompetitive business practices to fraud, false advertising, and, now, dark patterns.[80] Describing the evils that the FTC would confront, Congress provided simply that "[u]nfair methods of competition . . . and unfair or deceptive acts or practices in commerce . . . are hereby declared unlawful."[81] That is it—no list of unscrupulous practices or safe harbors for scrupulous ones, just a broad prohibition on "unfairness" and "deception" in the market.[82] Nor was the broad terminology, as is sometimes the case, later limited by more precise regulations. Even today, more than 100 years later, the FTC's own definitions of "unfair" and "deceptive" remain nearly as broad as Congress's: According to the FTC, a "deceptive" act can be any material misrepresentation likely to mislead a reasonable consumer,[83] and an "unfair" act any practice

---

[78] Thus, in addition to prohibitions on specific techniques, California's privacy regulations also bar interfaces that impose "unnecessary burden or friction on users." CAL. CODE REGS. tit. 11 § 7004(a)(5).

[79] *See generally* Federal Trade Commission Act, Ch. 311, 38 Stat. 717 (1914) (codified as amended at 15 U.S.C. §§ 41–58) (creating the FTC and defining its powers and duties).

[80] *See* 15 U.S.C. §§ 45–54 (empowering the FTC to regulate competition, fraud, false advertising, and other unfair business practices).

[81] § 45(a)(1).

[82] *See* § 44 (leaving undefined the terms "unfair" or "deceptive" acts).

[83] FTC Policy Statement on Deception, appended to Cliffdale Associates, Inc., 103 F.T.C. 110, 174 (1984) from FTC Chairman James C. Miller III to Committee on Energy and Commerce Chairman John D. Dingell (October 14, 1983) [hereinafter FTC Policy Statement on Deception], https://www.ftc.gov/system/files/documents/public_statements/410531/

that causes unjustified consumer injury in violation of public policy.[84]

Even with the advantage of hindsight from decades of enforcement actions, the FTC has added little detail to what constitute unfair or deceptive practice beyond the statutory text, or, for that matter, the similar common-law doctrines of fraud and unfair competition that preceded it. The point is this: Lawmakers can and do offer guidance documents that explore how broad concepts of unfairness and deception might apply in particular, modern contexts,[85] but a precise ex ante definition is impossible.[86] An attempt to write into law a precise boundary between persuasion and coercion is certain to be incomplete, inaccurate, imprecise, or, very possibly, all three.

*3. Accidental Deception.* A third challenge facing lawmakers is the difficulty of designing modern user interfaces.[87] Major tech companies devote entire teams to tweaking and optimizing user interfaces for ease of use, legal compliance, and targeted ads. But interface design is difficult and expensive.[88] For smaller companies

831014deceptionstmt.pdf ("The Commission believes that to be deceptive the representation omission, or practice must be likely to mislead reasonable consumers under the circumstances.").

[84] FTC Policy Statement on Unfairness, appended to International Harvester Co., 104 F.T.C. 949, 1070 (1984) from FTC Commissioners Michael Pertschuk, Paul Rand Dixon, David A. Clanton, Robert Pitofsky & Patricia P. Bailey to Consumer Subcommittee members Wendell H. Ford & John C. Danforth (December 17, 1980) [hereinafter FTC Policy Statement on Unfairness], https://www.ftc.gov/legal-library/browse/ftc-policy-statement-unfairness (defining unfair acts as those injurious to consumers, violative of public policy, unethical, or unscrupulous).

[85] *See generally* BRINGING DARK PATTERNS TO LIGHT, *supra* note 40 (summarizing an FTC public workshop on digital dark patterns and highlighting recommendations for companies to adhere to in light of new developments in digital marketing).

[86] *See, e.g., id.* at 3 (noting the evolving nature of unfair and deceptive practices due to new technologies such as augmented reality and virtual reality, which makes an ex ante definition to "unfair" and "deceptive" ineffective).

[87] *See* Hurwitz, *supra* note 34, at 77 ("In some systems, including nearly all software-based systems, design is more than just difficult, it is 'complicated.' Complex systems are systems with many interconnected parts, in which changes to any one of those part can affect other parts, often in unexpected and hard to understand ways.").

[88] *See* Kimberlee Leonard, *How Much Does a Website Cost? (2023 Guide)*, FORBES: ADVISOR (Jan. 15, 2023, 9:23 PM), https://www.forbes.com/advisor/business/software/how-much-does-a-website-cost/#website_costs_by_industry_section (outlining the many expenses involved in designing a website for businesses of all sizes).

on limited budgets, top-flight designers and compliance experts may be out of reach. In such cases, complex, confusing user interfaces may be the product of inexperience and underfunding rather than an intent to deceive.[89] Overly burdensome interface design restrictions could impede product innovation by increasing development costs.

A related challenge is what to do about A/B testing of interface designs. When paired with machine-learning-driven optimization, A/B testing takes humans out of the design equation altogether. Is a pattern still "dark" if it is accidentally, rather than intentionally, deceptive? Historically the FTC's approach has been to focus on the *effect* of deceptive conduct, not the intent behind it, prohibiting, for example, advertisements that confuse a "substantial minority"[90] of consumers, even if unintentionally. But perhaps the standard should be more lenient for app and website designs than for traditional advertisements given the comparative difficulty of designing multifunction user interfaces.

*4. Limited Enforcement Resources.* Lawmakers also face a more practical obstacle: limited enforcement resources. Last year alone, the FTC received more than five million reports of fraud. Of those, 300,000 related to online shopping and another 300,000 to internet services, telephone services, data privacy, electronic media, and computer equipment and software—categories that include typical dark-pattern techniques such undisclosed costs, difficulty canceling online accounts, unauthorized charges in mobile apps, use of deceptive practices to obtain consumer data, online streaming, and unauthorized installations and downloads.[91] And those numbers are increasing. Complaints related to online shopping roughly doubled

---

[89] *See* Hurwitz, *supra* note 34, at 77–90 (noting the difficulty and costs of user interface design and urging caution before concluding that suboptimal interfaces are malicious).

[90] *See* FTC Policy Statement on Deception, *supra* note 83, at 2 n.20 (providing that "[a] material practice that misleads a significant minority of reasonable consumers is deceptive"); FTC Enforcement Policy Statement on Deceptively Formatted Advertisements, 81 Fed. Reg. 22596 (Apr. 18, 2016) (same).

[91] *See FTC Consumer Sentinel Network*, TABLEAU PUB. (Feb. 23, 2023), https://public.tableau.com/app/profile/federal.trade.commission/viz/TheBigViewAllSentinelReports/CategoriesRanked (reporting by category consumer complaints to the FTC).

from 2018 to 2022[92] and may remain elevated indefinitely as consumers embrace online shopping.

Handling this volume would be challenging enough were combating dark patterns and online fraud the FTC's sole mission. But its responsibilities are far broader than that. Its mission also includes protecting the public from identity theft, data breaches, anticompetitive mergers and acquisitions, and a host of other dangers.[93] Yet, despite its broad mandate, the FTC receives less funding and employs a leaner staff than its international counterparts.[94] In 2022 the FTC employed approximately 1,100 full-time employees, a number that has remained constant over the last three decades.[95] Indeed, the FTC's current staffing "is 600 fewer full-time staff [than at] the beginning of the Reagan Administration."[96]

All of this means that even if lawmakers were to craft new rules targeting dark-pattern trickery, agency resources would continue to present a major constraint on enforcement. Much of the troublesome behavior in online marketplaces persists not because it is currently lawful but because resources to track down violators are so limited. In the FTC's case, for example, it initiated approximately 200 consumer-protection actions in all of 2022, the vast majority of which were unrelated to dark patterns.[97] State FTC analogues and the newly created California Privacy Protection Agency face similar staffing limitations. With enforcement resources so limited, even a perfectly designed dark-pattern law could be enforced against only the most egregious violators.

---

[92] *See id.* (showing that the FTC Consumer Sentinel Network had 154,084 complaints in 2018 and 327,609 in 2022).

[93] *Mission*, FED. TRADE COMM'N, https://www.ftc.gov/about-ftc/mission (last visited June 5, 2023).

[94] Transforming the FTC: Legislation to Modernize Consumer Protection Before the H. Subcommittee on Consumer Protection and Commerce, 117th Cong. 1 (2021) (Letter from Committee on Energy and Commerce Staff to Subcommittee on Consumer Protection and Commerce Members and Staff) [hereinafter Transforming the FTC], *available at* https://democrats-energycommerce.house.gov/sites/democrats.energycommerce.house.gov/files/documents/Briefing%20Memo_CPC%20Hearing_2021.07.28.pdf.

[95] *FTC Appropriation and Full-Time Equivalent (FTE) History*, FED. TRADE COMM'N, https://www.ftc.gov/about-ftc/bureaus-offices/office-executive-director/financial-management-office/ftc-appropriation (last visited June 5, 2023).

[96] Transforming the FTC, *supra* note 94; *see also* Testimony of Chair Lina M. Khan Before the House Appropriations Subcommittee on Financial Services and General Government at 3 (May 18, 2022) (observing that the FTC's "total headcount today remains about two-thirds of what it was at the beginning of 1980").

## IV. Why Private Law for Dark Patterns?

Most discussion of dark patterns to date has focused on public law: whether the proposed DETOUR Act or some other fresh legislation is necessary, whether antitrust law should be recalibrated, or whether the FTC Act's prohibition on unfair or deceptive practices sufficiently addresses dark patterns already.[98] Missing from the conversation, however, has been private law—that collection of legal doctrines such as property, tort, restitution, and contract law that govern horizontal relationships between private persons rather than the vertical relationship between individuals and the government.

Private law's absence from the conversation is striking, as its features would seem to make it especially well suited to the task. This Part explores those features of private law that may give it an advantage over public law for counteracting dark patterns, sketches a few possible courses of action, and concludes by exhorting scholars, courts, and legislatures to consider a role for private law in policing dark patterns.

### A. DECISIONAL LAW'S FLEXIBILITY

First, private law is typically developed by courts, as they adjudicate individual disputes on a case-by-case basis, after the incident in question, and with all available facts. Consider the tort of negligence, which imposes liability for causing physical harm to another through one's failure to "exercise reasonable care under all the circumstances."[99] Unlike a legislature establishing a forward-looking rule of conduct, a court determining whether a defendant's conduct amounts to negligence need not imagine beforehand all conceivable human conduct and categorize it as either reasonable or unreasonable. Instead, looking backward and considering all of the circumstances, a court decides whether the defendant's acts were consistent with the behavior of a reasonably prudent person.[100] As time passes and more cases are decided, trends in decisions will emerge and form rule-like precedents that can guide future decisions. But those precedential rules are created ex post. Legislatures are never forced to define the full scope of the law ex ante, an impossible task for dark patterns.

---

[97] *See Legal Library: Cases and Proceedings*, Fed. Trade Comm'n, https://www.ftc.gov/legal-library/browse/cases-proceedings (last visited June 5, 2023) (supplying a database of FTC actions).

[98] *See supra* sections II.A–B.

[99] Restatement (Third) of Torts: Liab. for Econ. Harm § 3 (Am. L. Inst. 2020).

A related benefit of private law is its power to assimilate the collective wisdom of judges around the country. For all the consternation it causes to law students and civil lawyers, there is a beautiful method to the common law's madness: As society learns, so do its lawyers, its judges, and its law. The process is incremental. New technologies present new cases, and the common law stretches, creating new lines of precedent to accommodate changing times. Where a strand of case law breaks down under the strain of societal developments, showing itself to be no longer suitable to its purpose, it is cut off and cast aside so that the law can move in a new direction. Because the particulars of private law are generated from court decisions, the law is able to adjust as society and technology evolve, just the sort of flexibility that dark patterns require.

At least that is the hope. The common law approach has its drawbacks. Precedent-based decision making sacrifices ex ante certainty for ex post flexibility and provides less certainty to regulated parties.[101] The approach shines most brightly in those contexts where ex ante rules are impossible to craft anyway, and so there is little loss of certainty to regulated parties by delaying classification of behavior as lawful or unlawful until after all facts are in hand.

---

[100] *See* Giacomo A. M. Ponzetto & Patricio A. Fernandez, *Case Law vs. Statute Law: An Evolutionary Comparison*, 37 J.L. STUD. 379, 379 (2008) ("Case law develops gradually through the rulings of appellate judges who have heterogenous preferences but are partially bound by stare decisis. We show that its evolution converges toward more efficient and predictable legal rules. Since statutes do not share this evolutionary property, case law is the best system when the efficient rule is time invariant, even if the legislature is more democratically representative than individual judges are."). Of course, regulatory enforcement actions form decisional law too. Indeed, that is how the FTC has generally policed this area. That body of decisional law guides both regulated entities and future enforcement actions, but given the FTC's limited enforcement resources, it is less robust and evolves more slowly than might judicial decisional law. Regardless, the bigger point here is that decisional law responding to dark patterns is superior to attempts by legislation or regulation to ban specific techniques or technologies.

As it happens, dark patterns are just such a context. Lawmakers targeting dark patterns have found it necessary to resort to broad, catch-all language that offers little advantage over judge-made decisional law.[102] And for good reason. Commercial shenanigans are as old as humanity. Dark patterns are just a recent, quickly evolving example. This makes them perfect for resolution through precedent-driven, judicial lawmaking. Judges empowered with the broad mandates like those of tort and contract law to root out "material misrepresentations," "justifiable reliances," and "fraudulent inducements" would have exactly the tools they need—a double measure of wit and a boatload of human experience—to sniff out abusive dark patterns in whatever shape they appear. As dark patterns evolve, so too could the body of law that tracks them.

## B. PRIVATE ENFORCEMENT RESOURCES

A second benefit of private law is its diffusion of enforcement authority to the public, whose members vastly outnumber administrative officials and are often better situated to identify unlawful behavior. A private law approach to dark-pattern regulation would effectively deputize a nation of "private attorneys general" that could police dark patterns far more thoroughly than can the leanly staffed FTC.

Fraud claims are a useful example. Common-law fraud occurs where a defendant makes a material misrepresentation that is intended to and does induce another to act to her economic detriment.[103] Given the requirement for a "material

---

[101] For detailed discussion of this concept, see generally Louis Kaplow, *Rules Versus Standards: An Economic Analysis*, 42 DUKE L.J. 557 (1992); Pierre J. Schlag, *Rules and Standards*, 33 UCLA L. REV. 379 (1985); Gideon Parchomovsky & Alex Stein, *Catalogs*, 115 COLUM. L. REV. 165, 172–81 (2015).

[102] *See supra* sections II.A–B.

[103] *See* RESTATEMENT (THIRD) OF TORTS: LIAB. FOR ECON. HARM § 9 (AM. L. INST. 2020) (providing that "[o]ne who fraudulently makes a material misrepresentation of fact, opinion,

misrepresentation" that "induces another to act" in "justifiable reliance," common-law fraud claims almost invariably also constitute deceptive trade practices in violation of the FTC Act. Yet private parties bring many more actions against fraudsters each year than the FTC could bring. The reason is simple. Private parties who have been defrauded have intimate knowledge of the conduct in question; they have every incentive to seek recovery of their losses; and private lawyers who can seek recovery on their behalf far outnumber the attorneys at the FTC charged with policing the competitive landscape.

This is the great benefit of private ordering. Individual citizens are often better informed of the circumstances of their losses, hold a greater interest in their own personal welfare, and, above all, are far more numerous than government officials. Private actions to protect property and contract rights have long been the norm, as have private actions in tort and restitution to vindicate personal interests against physical, economic, and emotional harms. And in modern times, legislatures have expanded the law to protect additional rights by creating private causes of action for, among other things, race- or sex-based discrimination in employment and housing, discrimination against persons with disabilities by commercial facilities and places of public accommodation, and the mishandling of an individual's personal information. The federal government even relies on private claimants to help it police fraud by allowing individuals to bring qui tam lawsuits on behalf of the U.S. under the False Claims Act and, if successful, to retain a portion of any award or settlement.

But private enforcement does not work in every context. Sometimes the harms of dark patterns are minor or noneconomic— privacy invasions, misuse of consumer data, or wasted time, for example. Individuals then have little incentive to sue, because their losses are small or nonexistent, and attorneys have little incentive to represent them because the scant recovery would not support the litigation expenses they would incur to prosecute the claim. Legislatures might overcome this obstacle by providing for attorney's fees, statutory damages, and other incentives to

---

intention, or law, for the purpose of inducing another to act or refrain from acting, is subject to liability for economic loss caused by the other's justifiable reliance on the misrepresentation").

encourage more enforcement, but such schemes must be carefully tailored to avoid the risk of overenforcement and overdeterrence, which could be a net negative to society.

In short, private enforcement is no panacea, but it works well for many problems, and dark patterns may be among them. The internet is a vast sea, brimming with small and nimble digital ne'er-do-wells. Dispersed policing by private parties could be part of the solution.

## C. A SMORGASBORD OF STRATEGIES

The primary point of this Article is to show why dark patterns present such a difficult problem for lawmakers and how private law is well suited to form part of the solution. The natural follow-up question is how to do it. What exactly should lawmakers do to focus the power of private law on dark patterns? A complete answer is beyond the scope of this Article but is an important area for future work. To guide those efforts, this final section sketches a few preliminary thoughts on possible strategies and pitfalls for lawmakers to avoid.

First, the good news: Applying private law to dark patterns does not require a huge departure from existing law. Many existing contract and tort-law principles are already flexible enough to respond to dark patterns. Where a dark pattern is confusing or coercive enough to undermine a consumer's consent, courts could refuse to enforce the agreement on the ground that the contract was fraudulently induced, that fraud prevented valid formation, or that the fraud justifies rescission of the contract. Courts could even enforce the contracts according to the terms as understood by the consumer by reforming the agreement or enforcing the deceptive expression as an implied contractual warranty. Alternatively, where no contract is present, the law might offer recovery via actions for conversion, fraud, breach of fiduciary duty, restitution, deceptive marketing, or simple negligence. Centuries of judicial evolution have produced extensive bodies of contract, tort, and restitution law with endlessly flexible tools to address all manner of lies and the lying liars who tell them.

Despite their flexibility, however, private law has been slow out of the gate. Courts have been cautious, perhaps overly so, when expanding old legal principles to the dark-pattern context. And, given the typically small losses suffered by dark-pattern victims and the difficulty identifying defendants, it often makes little sense for a customer to litigate a claim. That is especially so for those dark-pattern victims who face no monetary loss at all, just the annoyance and inconvenience of having their time wasted as they uncheck unwanted defaults, unsubscribe from mailing lists they never intended to join, or discontinue a transaction after learning at the end of the checkout process about the seller's 200% shipping and handling fee. Thus, although private law offers the benefit of flexibility, harnessing its power may require legislative intervention to create a pathway for litigants.

Fortunately, the problem is a familiar one that has been surmounted in other contexts. For example, faced with an avalanche of telemarketing robocalls, Congress enacted the Telephone Consumer Protection Act of 1991 (TCPA), which provides for statutory damages of $500 per call for plaintiffs who successfully prosecute claims against companies who use automated telephone dialing systems to contact individuals without their consent. Similarly, when it enacted the Americans with Disabilities Act and the Fair Housing Act, Congress intentionally leveraged the on-the-ground knowledge of individuals and the enforcement resources of private attorneys by allowing for the recovery of attorneys' fees to successful litigants of claims brought pursuant to these statutes. The same sort of approach could work for dark patterns. A law permitting awards of treble damages, or providing minimum statutory damages or attorney's fees, could motivate plaintiffs to bring claims challenging dark patterns even in cases where their actual harms would not otherwise support litigation.

Second, this approach also has its dangers. One risk is overenforcement. Supercompensatory damages spur private enforcement, but they do so by offering damages beyond consumers' actual economic harms. Legislators could easily overshoot and set damages too high, which would encourage actions for de minimis violations or quick settlements of meritless claims. The result would be increased litigation risk for often-small tech companies that could discourage innovative apps and interface designs. Another risk is balkanization of the law. Increased enforcement means more lawsuits in more jurisdictions. More lawsuits discourage misconduct and press the law to evolve more quickly to account for emerging dark patterns. But if the law governing dark patterns diverges across jurisdictions, website and app designers will face increase compliance costs to track its movements.

Legislators have tools available to account for these risks. Capping damage awards or limiting damages to some multiple of actual consumer economic harm could discourage overenforcement, as could a two-way fee provision that allows successful defendants to recover their own attorney's fees in the case of meritless claims. Or, as California has done with the CCPA, legislators might require notice to a prospective defendant and an opportunity to cure before a private party may commence litigation. To combat balkanization, legislators might consider a preemptive federal law or a federal safe harbor for established best practices. The possibilities are endless.

My point here is a general one—not to push for any particular approach, but to emphasize that careful planning will be required to craft a well-balanced incentive system. I hope to take up that important work in a future Article. Here, the aim is to explain the special difficulty that dark patterns present and to highlight the advantages that a private law approach can offer. Shifting enforcement to the private sphere would effectively deputize every American, and her attorney, to pursue dark-pattern abuses, dramatically expanding the enforcement resources of the FTC and state analogues.

## V. Conclusion

Lawmakers announcing the threat of dark patterns are half right. Dark patterns are a growing problem. They undermine human autonomy, bias customers' evaluation and selection of products, and distort online markets for goods and services. But new legislation is not the solution—at least, not the sorts of legislation on the table today. Existing law already bars the most problematic dark patterns, and enforcement resources are so limited that more detailed restrictions are unlikely to make a difference. Moreover, statutes and regulations calling out specific strategies and techniques are destined always to be one step behind the next technological evolution. Lawmakers should resist the temptation to target dark patterns directly and instead consider ways to leverage the power of private law. Encouraging private actions to challenge dark patterns will direct fresh resources to police dark pattern abuses and dramatically accelerate the pace at which the law of dark patterns evolves, pressing it to track new technologies of deception wherever they may lead.